\documentclass[12pt]{article}
\newcommand{\ba}{\begin{array}}
\newcommand{\ea}{\end{array}}

\newcommand{\be}{\begin{equation}}
\newcommand{\ee}{\end{equation}}
\newcommand{\bea}{\begin{eqnarray}}
\newcommand{\eea}{\end{eqnarray}}
\newcommand{\beal}{\setcounter{letter}{1} \begin{eqnarray}}
\newcommand{\eeal}{\addtocounter{equation}{1} \end{eqnarray}}
\newcommand{\none}{\nonumber \\}

\newcommand{\larrow}{\,\,\,\,\hbox to 30pt{\rightarrowfill}
\,\,\,\,}
\newcommand{\slarrow}{\,\,\,\hbox to 20pt{\rightarrowfill}
\,\,\,}

\newcommand{\bm}{\bibitem}
\begin{document}

\begin{titlepage}
\renewcommand{\thefootnote}{\fnsymbol{footnote}}
\renewcommand{\baselinestretch}{1.3}
\medskip
\hfill  UNB Technical Report 04-04\\[20pt]

\begin{center}
{\large {\bf Ricci Flow of 3-D Manifolds with One Killing Vector }}
\\ \medskip  {}
\medskip

\renewcommand{\baselinestretch}{1}
{\bf
J. Gegenberg $\dagger$
G. Kunstatter $\sharp$
\\}
\vspace*{0.50cm}
{\sl
$\dagger$ Dept. of Mathematics and Statistics and Department of Physics,
University of New Brunswick\\
Fredericton, New Brunswick, Canada  E3B 5A3\\
{[e-mail: lenin@math.unb.ca]}\\ [5pt]
}
{\sl
$\sharp$ Dept. of Physics and Winnipeg Institute of
Theoretical Physics, University of Winnipeg\\
Winnipeg, Manitoba, Canada R3B 2E9\\
{[e-mail: gabor@theory.uwinnipeg.ca]}\\[5pt]
}

\end{center}

\renewcommand{\baselinestretch}{1}

\begin{center}
{\bf Abstract}
\end{center}
{\small
We implement a suggestion by Bakas and consider the Ricci flow of 3-d manifolds with one Killing vector by 
dimensional reduction to the corresponding flow of a 2-d manifold plus scalar (dilaton) field. By suitably modifying the flow equations in order to make
them manifestly parabolic, we are able to show that the equations for the 2-d geometry can be put in the form explicitly solved
by Bakas using a continual analogue of the Toda field equations. The only remaining equation, namely that of the
scale factor of the extra dimension, is a linear equation that can be readily solved using standard techniques once the 2-geometry is specified.
We illustrate the method with a couple of specific examples. }
\vfill
\hfill  September 2004 \\
\end{titlepage}

\section{Introduction}
Ricci flow of 2-D manifolds is interesting because of its relationship
to the renormalization group equations of generalized 2-D sigma models
and because it provides a proof of the uniformization theorem in 2-D.
The uniformization theorem in two dimensions \cite{poincare} states
that every closed orientable
two dimensional manifold with handle number 0,1, or $>1$, admits
{\it uniquely}
the constant curvature geometry with positive, zero,
or negative curvatures, respectively.  

Recently, Bakas \cite{bakas} has shown that the 2-D Ricci flow equations in conformal gauge
provide a continual analogue of the Toda field equations. Using this algebraic approach
 he was able to write down the general solution.

The potential importance of a 3D uniformization theorem
is evident particularly in the context of
(super)membrane physics and three-dimensional quantum gravity
where one should be able to perform path-integral quantization via a similar
procedure to that in two dimensions.
Unfortunately, there is no uniformization theorem in three dimensions, only
a conjecture due to W.P. Thurston.
\cite{thurston,scott}.  

Recently there has been speculation that Perelman
\cite{perelman} has overcome some roadblocks in Hamilton's
program to prove the conjecture using the `Ricci flow'
\cite{hamilton1,caochow}.   It is therefore important to understand in
detail the properties of this flow.

In the following, we follow up on a suggestion by Bakas to use his 2-D results
in order to analyze the flow equations for 3-D manifolds with a single Killing vector.
This provides a tractable midisuperspace approach. We will show that the suitably modified flow
equations reduce to the infinite dimensional generalization of the 
Toda equation for the conformal factor of the invariant 2-d
submanifold plus a linear equation for the scale factor of the extra dimension.
 Note that since the latter scale factor depends on the coordinates of the invariant
 subspace, our manifolds are not simple direct products. We will analyze two
 exact analytic solution in detail and show that it has the expected behaviour.

The paper is organized as follows. Section II reviews 2-D flow equations and
Bakas' results, Section III
 presents the flow equations in the chosen coordinate system. It shows as well how to modify the equations to make them manifestly parabolic. The resulting 
 equations have a  particularly simple form. Section IV presents specific
solutions and Section V ends with conclusions and prospects for future work.

\section{The 2D Case}

We now summarize the methodology and results of Bakas\cite{bakas} since they play a crucial
role in the following. The Ricci flow equations, for arbirary 2-metric
$g_{AB}$ are:
\be
{\partial g_{ij}\over \partial t} = -R_{ij} + \nabla_i \xi_j + \nabla_j \xi_i.
\label{2d flow}
\ee
The last two terms (the so-called ``De Turck'' terms) incorporate the effects of all possible diffeomorphisms and can be chosen 
arbitrarily in order to simplify the equations and/or optimize convergence. 
Originally, De Turck \cite{detruck} chose the vector field 
\be
\xi^i:=g^{jk}\left(\Gamma^i_{jk}-\Delta^i_{jk}\right),
\ee
where $\Gamma^i_{jk}$ is the Christoffel connection with respect to the 
Riemannian metric $g_{ij}$ and $\Delta^i_{jk}$ is a fixed 
`background connection'.  The purpose was to replace the Ricci flow, which 
is only weakly parabolic, by an equivalent flow which is strongly parabolic.

Bakas chose to work in the conformal gauge:
\be
ds^2 = g_{ij}dx^idx^j= {1\over2}\exp(\Phi)(dX^2+dY^2).
\ee
In this gauge, without the need to add De Turck terms takes the form of a non-linear ``heat equation'':
\be
{\partial\over\partial t}e^\Phi = \nabla^2\Phi.
\label{2d heat equation}
\ee
 
The Toda equations describe the integrable interactions of a collection of two dimensional
fields $\Phi_i(X,Y)$ coupled via the Cartan matrix $K_{ij}$:
\be
\sum_jK_{ij}e^{\Phi_j(X,Y)}= \nabla^2\Phi_i(X,Y).
\ee 
Bakas argues that Eq.(\ref{2d heat equation}) is a continual analogue of the above,
with the Cartan matrix replaced by the kernel:
\be
K_{ij}\to K(t,t')={\partial\over\partial t}\delta(t,t').
\ee
This leads to a general solution to (\ref{2d heat equation}) in terms of a power series
around the free field expanded in path ordered exponentials. Although the resulting expression
is difficult to work with explicitly, it does provide a formal complete solution to the 2d flow
equations.

In the next section we will show that a similar formal solution also applies to the flow equations for 
three dimensional metrics with at least one Killing field.

\section{The 3D Case with One Killing Vector Field}
We consider the case where there the metric admits at least one Killing vector
field\footnote{This is the `midisuperspace flow' examined by Isenberg and Jackson\cite{isen}}.  We choose coordinates $(w,x^B)$ such that $\partial/\partial w$ is the
Killing vector field.  The upper case latin indices take the values $1,2$.  The metric can
be written in the form
\be
ds^2=e^\sigma(dw^2+A_B dx^B)^2+e^\phi \delta_{BC} dx^B dx^C.
\ee
The functions $\sigma,A_B,\phi$ depend only on the coordinates $x^B$.  

In this coordinate system, the naive Ricci flow equations for the metric:
\be
\dot g_{ij} = -2 R_{ij},
\ee
do not preserve the diagonal form of the metric, nor are they manifestly parabolic. We therefore considera metric flow equation suggested by the De Turck modifications:
\be
\dot g_{ij} = -2 R_{ij}-e^{-\sigma/2}L_\xi g_{ij}.
\ee
The vector field $\xi^m:=g^{mn}\partial_n e^{\sigma/2}$.  This can be written entirely in terms
tensors since
\be
\left|{\partial\over\partial w}\right|^2=e^\sigma.
\ee
Note, however, that unless $\sigma$ is constant, the extra term does not correspond precisely to a diffeomorphism of the metric. 

Using MAPLE/GRtensor, it follows that the explicit form of the flow for the metric above
is
\bea
\partial_t e^\phi &=& \Delta\phi +e^{\sigma-\phi}F,\\
\partial_t F &=& e^{-\sigma}\left[\Delta \left(e^{\sigma-\phi} F\right)-\nabla\sigma
\cdot \nabla\left(e^{\sigma-\phi}F\right)\right],\\
\partial_t e^\sigma &=& e^{\sigma-\phi}\Delta\sigma-\left(e^{\sigma-\phi}F\right)^2
\eea
In the above, all vector operators are with respect to the 2D Euclidean metric.  The
quantity $F:=\partial_1 A_2-\partial_2 A_1$ is the curl of the `twist potential'
$A_B$.  

We now examine the special case where the twist is trivial, that is,
$F= 0$. We also write $x=x^1,y=x^2$.   The flow reduces to the Toda flow for
$\phi$:
\be
\partial_t e^\phi=\Delta\phi
\ee
plus the flow
\be
e^\phi \partial_t \sigma=\Delta\sigma
\ee
The latter is not quite the linearized Toda equation.  But given a solution of the
Toda equation for $\phi$, we can in principle solve the second for $\sigma$.  

We look at two classes of solutions of the flow equations.
The first is obtained by choosing $\sigma=\phi$, so that the
metric is conformally flat, and the flow for the $\sigma$ field
is just another copy of the Toda equation.  

Hence for the sausage solution \cite{bakas} of the Toda equation:
\be
e^\phi={2\sinh{[2\gamma(t_0-t)]}\over \gamma(\cosh{[2\gamma(t_0- t)]}+\cosh{(2 y)})}.
\ee
In the limit as $t\to\infty$, $e^\phi\to 2/\gamma$, and the Ricci tensor 
goes to zero.  Hence, in this limit, the geometry is flat.  On the other 
hand, in the limit
$t\to t_0^+$, $e^\phi\to 2(t-t_0)/\cosh^2{y}$.  In the last limit, we find
that the Ricciscalar $R\sim\frac{1}{2(t-t_0)}(5-\cosh(2y))$.  So, if we 
flow the highly curved non-homogeneous metric 
\be
g^{(0)}(x,y,z;t=t_0+\epsilon):=\frac{2\epsilon}{\cosh^2(y)}\delta_{ij},
\ee
we end up at $t\to\infty$ with the flat metric.  This is consistent with 
Thurston's conjecture.

The second type of solution is of the Liouville type.  We set
\be
e^{\phi(x,y;t)}=T(t) e^{\psi(x,y)}
\ee
Now for $t\geq t_0$, we find that
\bea
T(t)&=&\beta(t-t_0)\none
\Delta\psi&-&\beta e^\psi=0,
\eea
where $\beta$ is a separation constant.  The second of the above 
equations is the Liouville equation, so the two dimensional part of the metric,
$e^\phi(dx^2+dy^2)$ has constant negative curvature (for $t\geq t_0$).  

One case has $\sigma=a \phi$ for a constant $a$.  In this case
\be
ds^2=e^{a\phi}dw^2+e^\phi(dx^2+dy^2).
\ee
The second case is that $\sigma=H(t)\rho(x,y)$. There are two solutions: 
\be
H(t)=\pm H_0(t-t_0)^{\gamma/\beta},
\ee
with the $+$ (respectively $-$) sign corresponding to positive (respectively negative) sign for the separation constant $\beta$.  
The function $\rho$ satisfies the linear partial differential equation
\be
\Delta\rho-\gamma e^\psi\rho=0.
\ee
Hence the metric is
\be
ds^2=e^{\pm H_0(t-t_0)^{\gamma/\beta}\rho}dw^2\pm\beta(t-t_0)e^\psi(dx^2+dy^2).\label{lioumetric}
\ee
The quantity $\psi$ is a solution of the Liouvile equation.

Consider first the case $t\leq t_0$.  If $\gamma\geq\beta$, then the flow starts from 
some highly curved non-homogeneous metric as $t\to-\infty$.  As $t\to t_0^-$, we have 
\bea
R_{AB}&\sim&\frac{1}{2(t_0-t)}g_{AB},\none
R_{ww}&\sim& 0,
\eea
with $A,B,...=x,y$.  Hence, the geometry is asymptotically that of the 
homogeneous, but anisotropic geometry $S^2\times E^1$. 

Similarly, for the case of $t\geq t_0$, the flow is `backwards' from a 
non-homogeneous geometry for $t> t_0$ to the homogeneous geometry 
$H^2\times E^1$. 

Thus the flow is consistent with the Thurston conjecture.  

\section{Conclusions}

We have shown that the Ricci flow equations for 3d metrics with at least one
Killing vector can be integrated in precisely the same manner as the 2d equations,
providing a particular choice of De Turck term is made. It is interesting to speculate
whether these techniques could work for more general 3 metrics. 

Consider, without loss of generality, a diagonal metric
\be
ds^2=e^{\phi_1(x;t)}(dx^1)^2+e^{\phi_2(x;t)}(dx^2)^2+e^{\phi_3(x;t)}(dx^3)^2
\ee
where the functions $\phi_i(x;t)$ depend on all 3 coordinates $x^i$.
The resulting  bare Ricci flow is again not manifestly elliptic, and the equations 
have non-trivial off-diagonal terms on the RHS that make direct integration difficult. Since in three
dimensions any metric can be made diagonal with a suitable coordinate transformation, it is
reasonable to assume that there exists a modified flow: 
\be
\dot g_{ij}=-2\left([ R_{ij}(g)+\Phi\left(\nabla_i\xi_j+
\nabla_j\xi_i\right)\right],
\ee
where $\Phi$ is some scalar field.  The extra term would ensure that diagonal metrics evolve into diagonal metrics. That is, one should look for
 fields $(\Phi,\xi)$ that diagonalize the right hand side of the above for arbitrary diagonal metric
$g_{ij}$. We have as yet not succeeded in this, but if such a flow did exist, it
is possible that the resulting three flow equations for each of the three scale factors
 would take a form similar to what we have found above, albeit with non-trivial coupling. It may therefore
provide a basis for solving the 3-d flow equations in a more general setting.

\noindent
{\bf Acknowledgements}:  

We wish to thank the Perimeter Institute, where much 
of this work was done, for its 
hospitality and support. This work was supporte by the Natural Sciences and Engineering Research Council of Canada.

\end{document}